\documentclass[aip,pop,amsmath,amssymb,reprint,fixfloat,graphicx]{revtex4-1}
\usepackage{color,graphicx}

\begin{document}


\title{Evolution of field line helicity during magnetic reconnection} 

\author{A.~J.~B.~Russell}
\email[]{Email: arussell@maths.dundee.ac.uk}
\affiliation{ Division of Mathematics, University of Dundee, DD1 4HN, U.K.}
\author{A.~R.~Yeates}
\affiliation{Department of Mathematical Sciences, Durham University, DH1 3LE, U.K.}
\author{G. Hornig}
\author{A.~L.~Wilmot-Smith}
\affiliation{ Division of Mathematics, University of Dundee, DD1 4HN, U.K.}

\date{\today}

\begin{abstract}
  We investigate the evolution of field line helicity for magnetic fields 
  that connect two boundaries without null points,
  with emphasis on localized finite-B magnetic reconnection.  
  Total (relative) magnetic helicity is already recognized as an important 
  topological constraint on magnetohydrodynamic processes.
  Field line helicity offers further advantages because it preserves all topological information 
  and can distinguish between different magnetic fields with the same total helicity.  
  Magnetic reconnection changes field connectivity
  and field line helicity reflects these changes;
  the goal of this paper is to characterize that evolution.  
  We start by deriving the evolution equation for field line helicity and examining its terms, 
  also obtaining a simplified form for cases where dynamics are localized within the domain.  
  The main result, which we support using kinematic examples,
  is that during localized reconnection in a complex magnetic field,
  the evolution of field line helicity is dominated by a work-like term
  that is evaluated at the field line endpoints, 
  namely the scalar product of the generalized field line velocity and the vector potential.  
  Furthermore, the flux integral of this term over certain areas 
  is very small compared to the integral of the unsigned quantity, 
  which indicates that changes of field line helicity happen in a well-organized pairwise manner.  
  It follows that reconnection is very efficient at redistributing 
  helicity in complex magnetic fields
  despite having little effect on the total helicity.
\end{abstract}


\maketitle 


\section{Introduction}\label{sec:intro}

Magnetic helicity is a valuable concept in magnetohydrodynamics (MHD) 
that quantifies the linking, twisting and kinking of magnetic field lines \citep{1969Moffatt,1999Berger}.
It is highly conserved under a broad range of circumstances and it
therefore has many applications in both laboratory and astrophysical plasmas \citep{1999BrownGMS}.
More precisely, total magnetic helicity 
(the volume integral of $\vec{A}\cdot\vec{B}$ where $\vec{A}$ is the vector potential 
and $\vec{B}=\nabla\times\vec{A}$ is the magnetic field) 
is an ideal MHD invariant for magnetically closed domains \citep{1958Woltjer},
which is readily extended to magnetically open domains either
as total relative magnetic helicity \citep{1984BergerFields,1985FinnAntonsen}
or by an appropriate restriction of the gauge of the vector potential \citep{2014PriorYeates}.
Ideal invariance holds because ideal evolutions neither create nor destroy magnetic helicity
and are unable to transport helicity across the magnetic field.
In non-ideal MHD, exact conservation of magnetic helicity breaks down
but magnetic reconnection at high magnetic Reynolds number
nonetheless conserves total (relative) magnetic helicity very well
\citep{1974Taylor,1984Berger}.
Reconnection does, however,  redistribute helicity 
between field lines as magnetic connectivities change.
Thus, magnetic reconnection approximately conserves total magnetic helicity 
but may radically alter how that total is composed.

An extreme example of the redistribution of magnetic helicity  is Taylor relaxation.
Considering a reversed-field pinch, \citet{1974Taylor,1986Taylor} hypothesized that
turbulent magnetic reconnection allows an initial magnetic field to relax to the 
minimum energy state with the same total helicity,
which had previously been shown by \citet{1958Woltjer} to be a linear force-free field.
This assumes that total helicity is the only helicity constraint
and requires complete redistribution of magnetic helicity across the cross-section of the device.
Taylor relaxation was successful for reversed-field pinches and has since been investigated for other situations
e.g. for the solar corona with applications to coronal heating and microflares 
\citep{1984HeyvaertsPriest,1986BrowningPriest,1989Dixon}.
There are also  known examples where the end state
is a nonlinear force-free field \citep{1980Bhattacharjee,2000Amari,2011Pontin},
however redistribution of helicity, although less extensive, is a major feature of those cases as well.

Due to its conservation, magnetic helicity is of broad astrophysical interest, 
especially in scenarios involving magnetic reconnection.
For instance, the generation of magnetic fields by dynamo action 
is intrinsically related to the properties of magnetic helicity \citep{2009Brandenburg}.
To give a few more examples, helicity conservation has been invoked 
in magnetospheric physics to explain generation of twisted flux tubes during dayside reconnection 
and plasmoid formation in the magnetotail \citep{1989WrightBerger,1989SongLysak}.
In solar physics, magnetic helicity is injected into the corona by flux emergence and 
photospheric motions including differential rotation and shearing flows in active regions\citep{2009DemoulinPariat}.
Values and changes of magnetic helicity in active regions
have been linked to solar flares and coronal mass ejections 
\citep{2002Moon,2004NindosAndrews,2010Park}, 
helicity ``condensation'' is a candidate explanation for the formation of filament channels \citep{2013Antiochos},
expulsion of helicity from corona leads to the presence of twisted flux ropes in the heliosphere \citep{1994Rust}
and reconnection of flux tubes can be a source of torsional Alfv\'en waves \citep{1986ShibataUchida,1987Wright}.

In this paper we consider a refined measure of helicity: 
a helicity density which is assigned to each field line.
This ``field line helicity'' contains all available topological information and can therefore 
distinguish between magnetic fields with the same total helicity.
The primary aim is to investigate how this measure of helicity evolves in the broad regime between
ideal evolution (for which every field line has its own helicity invariant)
and Taylor relaxation (for which the only helicity invariant is total helicity).
The results characterize changes to the composition of magnetic helicity
and are expected to advance our understanding of 3D magnetic reconnection
across a broad variety of applications including turbulent magnetic relaxation.

This paper is organized as follows.
Section \ref{sec:prelim} describes the model, 
recaps the concept of field line helicity and discusses gauge considerations.
In Sec.~\ref{sec:equation}, the evolution equation for field line helicity 
is derived and its terms are examined.
We then focus on cases where dynamics are localized within the domain 
(Sec.~\ref{sec:localized}) and show that evolution of field line helicity 
for a given field line is dominated by a work-like term
which has a well-organized structure of pairs of positive and negative rates of change.
Kinematic examples in Sec.~\ref{sec:examples} confirm our analytic results.
The paper ends with a summary in Sec.~\ref{sec:conc}.

\section{Preliminaries}\label{sec:prelim}
\subsection{Model}\label{sec:model}
\begin{figure}
 \includegraphics{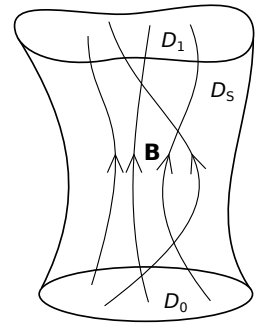}
 \caption{\label{fig:model}Sketch of the flux tube model with all field lines connecting $D_0$ to $D_1$.  
 The side of the domain, $D_S$, is a magnetic surface and the domain contains no magnetic nulls.}
\end{figure}

Much of what we discuss in this paper applies generally, 
but for concreteness we will consider a flux tube model sketched in Fig.~ \ref{fig:model}.
All field lines enter through a single surface $D_0$ and exit through a different surface $D_1$.
The remainder of the boundary is a magnetic surface, $D_S$, 
that joins the edge of $D_0$ with the edge of $D_1$.
There are no magnetic null points in the domain, 
hence we consider finite-B reconnection \citep{1988Schindler}.
This model can describe closed flux tubes 
(in which case magnetic field is periodic on $D_0$ and $D_1$) as well as open flux tubes.
More general magnetic fields can be partitioned into a collection of such domains,
which increases the generality of this model.

For simplicity, some of our results will be presented using a restricted version of the model.
The first simplification is to take $D_0$ and $D_1$ planar with 
outward surface normals, $\vec{n}$, pointing in the $\pm z$-direction.
In these cases, we also assume that the boundary is line-tied and ideal
so that electric field $\vec{E}=0$ and $\vec{B}\cdot\vec{n}$ is constant in time.
This restriction is appropriate to solar physics where the restricted model may, for example, 
represent the magnetic field in a coronal loop
under assumptions that the coronal magnetic field is evolving more rapidly 
than the photospheric convection timescale and that dynamics are concentrated away from the side boundary, $D_S$.
The simplified model is therefore of direct relevance to the relaxation of magnetic braids,
magnetic instabilities, coronal heating and confined solar flares.

\subsection{Field line helicity}\label{sec:helicity}

Field line helicity, $\mathcal{A}$, assigns a helicity value to every field line.
Physically, it measures the winding of magnetic flux  with 
the field line of interest\citep{2014PriorYeates} and
can also be viewed as a topological flux function
\citep{1988Berger,2011YeatesHornig,2013YeatesHornig,2014YeatesHornig}. 
It is defined as
\begin{align}\label{eq:curlyA}
  \mathcal{A}(\vec{x}) &= \int_{F(\vec{x})} \vec{A}\cdot \textrm{d}\vec{l}
\end{align}
where $\vec{x}$ is a point on a cross-section of our flux tube (e.g. the lower boundary, $D_0$),
$F(\vec{x})$ is the magnetic field line through that point
and $\textrm{d}\vec{l}$ is the line element along the field line.
Since our domain is free of null points the integral is always well defined.

Field-line helicity retains all topological information, 
in contrast to the volume-integrated total magnetic helicity\citep{2013YeatesHornig}.
It can therefore distinguish between topologically different magnetic fields with the same total helicity,
while the total magnetic helicity is easily recovered from $\mathcal{A}$ since 
\begin{align}\label{eq:total}
    H(\vec{A}) &= \int_V \vec{A}\cdot\vec{B} \ \textrm{d}^3x 
    = \int_{D_0} \mathcal{A} B_i \ \textrm{d}^2x 
    = \int_{D_1} \mathcal{A} B_o \ \textrm{d}^2x
\end{align}
where $B_i$ is the magnetic field component parallel to the inward normal on $D_0$
and $B_o$ is the magnetic field component parallel to the outward normal on $D_1$.
Note that this formula justifies the name field line helicity
since total helicity is the flux integral of $\mathcal{A}$.
Provided the gauge is suitably restricted, $\mathcal{A}$ has 
the desirable property of being an ideal invariant,
save for changes of field line connectivity caused by motions on the boundaries.
The gauge condition under which this is true is discussed in Sec.~\ref{sec:gauge} 
and the result is derived in Sec. \ref{sec:terms}.
At the same time, 
$\mathcal{A}$ is considerably easier to work with than the helicity density,
$\vec{A}\cdot\vec{B}$, which depends on all three coordinates 
and changes under ideal evolutions.

\subsection{Gauge considerations}\label{sec:gauge}

Magnetic helicity and field line helicity are in general gauge-dependent for open domains, 
i.e. they change under gauge transformations of the vector potential.
Referring to Eq.~(\ref{eq:curlyA}), 
a gauge transformation $\vec{A}^\prime=\vec{A}+\nabla \chi$ implies a new field line helicity
\begin{align}
  \mathcal{A}^\prime(\vec{x}) 
    &= \int_{F(\vec{x})} \left(\vec{A}+\nabla\chi\right)\cdot \textrm{d}\vec{l}
    &= \mathcal{A}(\vec{x})+\left[\chi\right]_{\vec{x}_0}^{\vec{x}_1},
\end{align}
where $\vec{x_0}$ and $\vec{x_1}$ represent the start and end points 
of the field line respectively.
It follows that field line helicity, although generally gauge-dependent, 
is invariant for a restricted set of gauge transformations
that have $\chi(\vec{x_0})=\chi(\vec{x_1})$. 
Therefore,  $\mathcal{A}$ may be made gauge invariant by imposing a 
boundary condition on the vector potential that
fixes the components of $\vec{A}$ tangent to the boundary.

The physical interpretation of $\mathcal{A}$
and the fixing of $\vec{n}\times\vec{A}$ on $\partial V$
can be regarded in two complementary ways. The first approach 
is to consider relative magnetic helicity\citep{1984BergerFields,1985FinnAntonsen},
\begin{align}\label{eq:relative}
  H_R(\vec{B}|\vec{B}^\textrm{ref}) &= 
      \int_V \left(\vec{A}+\vec{A}^\textrm{ref}\right)\cdot\left(\vec{B}-\vec{B}^\textrm{ref}\right) \ \textrm{d}^3x,
\end{align}
where ${\vec{B}^\textrm{ref}=\nabla\times\vec{A}^\textrm{ref}} $ 
is a reference field that satisfies 
$\vec{B}^\textrm{ref}\cdot\vec{n}|_{\partial V} = \vec{B}\cdot\vec{n}|_{\partial V}$.
It is common practice to choose the potential field for $\vec{B}^\textrm{ref}$.
Relative helicity has the advantage of being gauge invariant
but has its own disadvantage of depending on the chosen reference field,
which may need to change in time to match $\vec{B}(t)$ on $\partial V$.
If the gauge of $\vec{A}$ is restricted such that
\begin{align}\label{eq:gauge_condition}
  \vec{A}\times\vec{n}|_{\partial V}&=\vec{A}^\textrm{ref}\times\vec{n}|_{\partial V},
\end{align}
which is always possible given
${\vec{B}^\textrm{ref}\cdot\vec{n}|_{\partial V} = \vec{B}\cdot\vec{n}|_{\partial V}}$, 
then $\mathcal{A}$ becomes gauge invariant as noted above.
It can also be shown that Eq.~(\ref{eq:relative}) and (\ref{eq:gauge_condition}) imply that 
${H(\vec{B})=H_R(\vec{B}|\vec{B}^\textrm{ref})+H(\vec{B}^{ref})}$.
Therefore, if the boundary condition given by Eq.~(\ref{eq:gauge_condition})
is imposed using a reference field for which $H(\vec{B}^{ref})=0$, 
then $H=H_R$ and $\mathcal{A}$ can be regarded as the density per unit flux
of the gauge-independent relative helicity\citep{2013YeatesHornig}.

An alternative approach arises from work by \citet{2014PriorYeates}
who considered the physical interpretation of helicity
in open domains as measuring the winding between pairs of field lines.
This can be viewed as a generalization of the linking number 
interpretation of helicity in closed domains\citep{1969Moffatt}.
\citet{2014PriorYeates} concluded that each gauge measures winding with respect 
to a particular frame, but some gauges include a non-physical contribution 
equivalent to measuring winding in a twisted frame.
It is therefore desirable to restrict oneself to gauges that measure winding with 
respect to a fixed basis, which makes $\mathcal{A}$ gauge invariant for a given frame.
Furthermore, it has been shown that the choice of basis used to measure winding 
of field lines corresponds directly to a choice of reference field for 
relative helicity\citep{2014PriorYeates},
so there is an equivalence between the relative helicity and winding interpretations.

The general equations derived in Sec.~\ref{sec:equation} and \ref{sec:localized} 
are valid in any gauge of $\vec{A}$.
In the examples presented in Sec.~\ref{sec:examples},
we consider situations where the reference field is fixed in time 
(which is possible since $\vec{B}\cdot\vec{n}$ is time-invariant on $\partial V$). 
Furthermore, we choose a gauge such that $H=H_R$ 
with a potential reference field for $H_R$. 
Since these examples have uniform $\vec{B}\cdot\vec{n}$ on the boundary, 
this potential field is untwisted in the sense of \citet{2014PriorYeates}, 
meaning that $\mathcal{A}$ may be interpreted either as the average winding of field lines 
with respect to an untwisted basis, or as a density for the relative helicity. 
Since we have fixed the gauge of $\vec{A}$ on the boundaries over time, 
any change of $\mathcal{A}$ in time must correspond to a non-ideal process.

\section{Evolution equation}\label{sec:equation}

\subsection{Derivation}\label{sec:deriv}
The evolution equation for field line helicity can be derived as follows.
We start from a general form of Ohm's law, expressed as
\begin{align}\label{eq:ohms}
  \vec{E}+\vec{v}\times\vec{B} &= \vec{R},
\end{align}
where $\vec{R}$ is the non-ideal contribution to the electric field.
For any $\vec{R}$, the non-ideal term can be decomposed into a gradient term
that produces the parallel electric field,
plus a term perpendicular to $\vec{B}$ that sums with the gradient term
to give the correct perpendicular electric field, i.e.
\begin{align}\label{eq:R}
  \vec{R} &= \nabla\Psi - \vec{u}\times\vec{B}.
\end{align}
In this decomposition, $\Psi$ and $\vec{u}$ 
may not have simple analytic forms derivable from the local value of $\vec{R}$,
in fact they must generally be defined in a non-local manner.
Noting that Eq.~(\ref{eq:ohms}) and Eq.~(\ref{eq:R}) imply ${\vec{e}_B\cdot\nabla\Psi=E_{||}}$,
$\Psi$ is obtained from the integral
\begin{align}\label{eq:psi}
  \Psi(\vec{x}) &= \int_{F(\vec{x})} \vec{E}\cdot\textrm{d}\vec{l}.
\end{align}
Here, we consider situations without null points or closed field lines, 
hence $\Psi$ is well-defined and single-valued throughout the domain.
Having obtained $\Psi$ by integration of the parallel electric field,
the components of $\vec{u}$ perpendicular to $\vec{B}$
are readily computed from Eq.~(\ref{eq:R}),
completing the decomposition.

Evolution of the vector potential is governed by
\begin{align}\label{eq:faraday_A}
  \frac{\partial \vec{A}}{\partial t} &= -\vec{E}-\nabla\Phi
\end{align}
where $\Phi$ is the electric potential.
Using Eq.~(\ref{eq:ohms}) and (\ref{eq:R}) to substitute in (\ref{eq:faraday_A}),
\begin{align}\label{eq:faraday_A2}
  \frac{\partial \vec{A}}{\partial t} &= \vec{w}\times\vec{B}-\nabla\left(\Psi+\Phi\right),
\end{align}
where 
\begin{align}\label{eq:vel}
  \vec{w} &= \vec{v}+\vec{u}
\end{align}
is the generalized field line velocity.

The concept of generalized field line velocity has been discussed in detail by
\citet{2003HornigPriest} and \citet{HornigChapter}.
Briefly, taking the curl of Eq.~(\ref{eq:faraday_A2}) shows 
that the magnetic field evolves as if advected with $\vec{w}$.
The slip velocity $\vec{u}=\vec{w}-\vec{v}$ (and hence $\vec{w}$) 
generally depends on where one sets $\Psi=0$.
This nonuniqueness represents the fact that when magnetic connectivity is changing,
the generalized velocity of a field line depends on what plasma element it is traced from.
Another point worth noting is that
since $\vec{u}$ enters Eq.~(\ref{eq:R}) as $\vec{u}\times\vec{B}$,
the component $\vec{e}_B\cdot\vec{u}$ is not constrained
so one may choose a single component of $\vec{u}$ or equivalently $\vec{w}$.

We are now ready to derive an equation for
 $\textrm{d}\mathcal{A}/\textrm{d}t$.
Starting from the time derivative of Eq.~(\ref{eq:curlyA}),
we differentiate under the integral sign using the Lie-derivative 
formula for line integrals through 3D space\citep{1988Abraham,1997Hornig,2003Larsson},
then simplify using Eq.~(\ref{eq:faraday_A2}) to
obtain the line integral of a gradient, which is readily evaluated.  Thus,
\begin{align}\label{eq:dcurlyA}
  \frac{\textrm{d}\mathcal{A}(\vec{x},t)}{\textrm{d}t} 
  &= \frac{\textrm{d}}{\textrm{d}t}\int_{F(\vec{x},t)} \vec{A}\cdot \textrm{d}\vec{l} \nonumber\\
  &= \int_{F(\vec{x},t)} \left[ 
                \frac{\partial \vec{A}}{\partial t} 
                -\vec{w}\times\nabla\times\vec{A}
                +\nabla\left(\vec{w}\cdot\vec{A}\right)
        \right]\cdot \textrm{d}\vec{l} \nonumber \\
  &= \int_{F(\vec{x},t)} \nabla\left(\vec{w}\cdot\vec{A}-\Psi-\Phi \right) \cdot \textrm{d}\vec{l} \nonumber \\
  &= \left[ \vec{w}\cdot\vec{A} -\Psi-\Phi\right]_{\vec{x}_0}^{\vec{x}_1}
\end{align}
where $\vec{x_0}$ and $\vec{x_1}$ represent the start and end points of the field line respectively.

Thus, the evolution of field line helicity depends on 
the motion of field line end points parallel to the vector potential on the boundaries,
the integral of $E_{||}$ along the field line of interest (i.e. the voltage drop)
and the difference in electric scalar potential between the field line's end points.

\subsection{Interpretation of terms}\label{sec:terms}

The $\Phi$ terms in Eq.~(\ref{eq:dcurlyA}) correspond to 
a difference of scalar potential between the endpoints of the field line
and represent helicity flux along the field line due to the gauge.
It is the only term that remains in an ideal evolution with no motions on the boundaries.
It is usually convenient to use a gauge in which this term vanishes
so that helicity flux is due to physical terms only and $\mathcal{A}$ becomes an ideal invariant
save for changes of field line connectivity caused by motions on the boundaries.
As an example, in our restricted model with $\vec{E}=0$ on $\partial V$
one can impose a condition that
$\vec{A}\times\vec{n}$ on $\partial V$ is constant in time,
which makes $\Phi$ spatially constant on $\partial V$ by Eq.~(\ref{eq:faraday_A}),
hence the $\Phi$ terms cancel for every field line.
Physically, this restriction ensures that $\mathcal{A}$ measures winding 
with respect to a time-independent frame, eliminating non-physical changes
\citep{2014PriorYeates}, or equivalently that $\mathcal{A}$ is a field line helicity 
for relative helicity with a time-independent reference field\citep{2013YeatesHornig}.

The $\Psi$ terms correspond to a net voltage drop along the field line.
This quantity, $\int_{F}\vec{E}\cdot\textrm{d}\vec{l}$, 
has played prominent role in general magnetic reconnection.
Such a voltage drop across a localized non-ideal region is necessary and sufficient
for the change in connectivities to be felt outside that region, i.e. 
a net voltage drop across a localized non-ideal region distinguishes finite-B reconnection 
with global effects from finite-B reconnection with only local effects \citep{1988Schindler}.
Voltage drops are also widely used to measure reconnection,
with the maximum (unsigned) voltage drop quantifying the reconnection rate 
\citep{1975Vasyliunas,2005Hesse,1993HesseBirn}.

Lastly, the $\vec{w}\cdot\vec{A}$ terms represent motion 
of field line end points on the boundaries.
Their form is analogous to work done against a force
and depends only on the component of $\vec{w}$ parallel to the vector potential.
This term can be present for ideal evolutions when motions
on the boundaries change field line connectivity, 
in which case $\vec{w}\cdot\vec{A}=\vec{v}\cdot\vec{A}$.
It is also present during localized reconnection when $\vec{w}$ is due to field line slipping.
The terms are obviously gauge dependent since a change of $\vec{A}$ 
will in general change the pattern of $\vec{w}\cdot\vec{A}$ on $D_0$ and $D_1$
but they can be made gauge independent by fixing $\vec{A}\times\vec{n}$ on 
$\partial V$ (as recommended to make the $\Phi$ terms cancel for every field line)
and using freedom of the parallel component of $\vec{u}$ 
to set $\vec{w}\cdot\vec{n}=0$ on $D_0$ and $D_1$.
It will also be shown in Sec.~\ref{sec:pairing} that the contribution integrated 
over certain areas is independent of gauge even when 
no boundary condition is imposed on $\vec{A}$.
Removing this term entirely by gauge choice would require a time-dependent gauge 
on the boundary, which would remove the physical interpretation of $\mathcal{A}$
as measuring changes in field line connectivity within the domain.

\section{Localized reconnection}\label{sec:localized}
\subsection{Simplifications}
Several simplifications can be made when reconnection dynamics
 are localized within the domain.
For simplicity, we do this here for planar and horizontal boundaries $D_0$ and $D_1$.
We also assume that the electric field vanishes on the domain boundary 
since dynamics are internal.

First, it is convenient to use the freedom available in the definition of $\vec{u}$ 
to set $\vec{w}\cdot\vec{e}_z=0$ throughout the domain,
which closes $D_0$ and $D_1$ to transport of magnetic flux with $\vec{w}$.
Under this choice,
\begin{align}\label{eq:w_simplified}
  \vec{w} &= \frac{\vec{e}_z\times\left(\nabla\Psi-\vec{E}\right)}{B_z},
\end{align}
which may be confirmed by using Eq.~(\ref{eq:R}) 
to substitute for $\vec{R}$ in Eq.~(\ref{eq:ohms}),
then taking the cross product with $\vec{e}_z$ and rearranging for $\vec{w}$.
Since $\vec{E}$ is assumed zero on $\partial V$, 
the field line velocity on the boundaries is simply
\begin{align}\label{eq:w_boundaries}
  \vec{w} &= \frac{\vec{e}_z\times\nabla\Psi}{B_z},
\end{align}
which implies that field line end points move along contours of $\Psi$.

Next, we can choose to set $\Psi=0$ everywhere on $D_0$, 
which gives $\vec{w}=0$ there by Eq.~(\ref{eq:w_boundaries}).
We can also use a gauge condition that $\vec{n}\times\vec{A}$ is time-independent
to make the $\Phi$ terms vanish (Sec.~\ref{sec:terms}),
thereby ensuring that $\mathcal{A}$ is constant in the absence of reconnection.
Doing so, Eq.~(\ref{eq:dcurlyA}) simplifies to
\begin{align}\label{eq:dcurlyA_localized}
  \frac{\textrm{d}\mathcal{A}}{\textrm{d}t} 
  &= \vec{w}(\vec{x}_1)\cdot\vec{A}(\vec{x}_1)-\Psi(\vec{x}_1),
\end{align}
which contains only terms that are evaluated at a single boundary (the one where $\Psi\neq0$).

Finally, using Eq.~(\ref{eq:w_boundaries}) and noting that 
$\nabla\times\vec{e}_z=0$, on $D_0$ and $D_1$,
\begin{align}\label{eq:wdotA_boundary}
  \vec{w}\cdot\vec{A}
  &=-\frac{\vec{A}\cdot\nabla\times\left(\Psi\vec{e}_z\right)}{B_z}\nonumber\\
  &=\frac{\nabla\cdot\left(\vec{A}\times\left(\Psi\vec{e}_z\right)\right)}{B_z}-\frac{\Psi\vec{e}_z\cdot\nabla\times\vec{A}}{B_z}\nonumber\\
  &=\frac{\nabla\cdot\left(\Psi\vec{A}\times\vec{e}_z\right)}{B_z}-\Psi,
\end{align}
i.e. the work-like term can also be expressed as a divergence term 
minus the voltage drop along the field line.

\subsection{Dominance of work term}\label{sec:dominant}
The character of the evolution of field line helicity depends on which term dominates 
the right hand side of Eq.~(\ref{eq:dcurlyA_localized}).
Referring to Eq.~(\ref{eq:w_boundaries}), the work-like term scales as
\begin{align}
  \vec{w}\cdot\vec{A} &\sim \frac{L}{l}\Psi
\end{align}
where $L$ is the length scale associated with the vector potential on the boundary
($B_z=\vec{e}_z\cdot\nabla\times\vec{A}\sim A/L$)
and $l$ is the length scale associated with $\Psi$ on the boundary 
(${||\nabla{\Psi}||\sim\Psi/l}$).
For the purpose of this scaling argument, $A$ and $\Psi$ are characteristic values.
The nature of the evolution equation therefore depends on the ratio of length scales $L/l$,
with the work-like term dominating for $L/l\gg 1$.

The property $L/l\gg 1$ is characteristic of 
magnetic fields with complex field line mappings,
for example magnetic braids.
In these cases, the complexity of the field line mapping 
means that field line integrated quantities such as $\Psi$
vary on a length scale much smaller than the typical gradient length scale of 
quantities on the boundary that have not been integrated through the domain, 
including the vector potential \citep{2009WilmotSmith}.
This is because, for a sufficiently complex magnetic field, 
a pair of field lines traced from nearby starting points
may take very different paths through the domain,
thus acquiring very different contributions to the integrand.
For example, when the parallel electric field is integrated 
along field lines to obtain $\Psi$,
one field line may pass through a non-ideal region
that the other field line bypasses altogether.
In this way $l$ becomes much less than $L$.
We therefore conclude from the scaling analysis
that changes to a field line's value of $\mathcal{A}$ occur primarily through the work-like term
when magnetic reconnection occurs in a complex magnetic field.

\subsection{Paired increases and decreases}\label{sec:pairing}
Having considered the change of field line helicity for a chosen field line,
it is also instructive to look at changes integrated over area.
Integrating Eq.~(\ref{eq:wdotA_boundary}) over an area $S\subseteq D_1$ and
using the weight $B_z$ for consistency with the total helicity integral defined in Eq.~(\ref{eq:total}),
one finds
\begin{align}\label{eq:wdotA_int}
  \int_S B_z \vec{w}\cdot\vec{A} \ \textrm{d}^2x &=  
    \int_{\partial S} \Psi \vec{A}\times\vec{e}_z \cdot \vec{m} \ \textrm{d}x -\int_S B_z\Psi \ \textrm{d}^2x
\end{align}
where we have used the divergence theorem and 
$\vec{m}$ is an outward edge normal on $\partial S$.

There are certain choices of $S$ for which the first term on the right hand side of Eq.~(\ref{eq:wdotA_int}) vanishes
to leave 
\begin{align}\label{eq:area_integrals}
  \int_{S} B_z \vec{w}\cdot\vec{A} \ \textrm{d}^2x &=-\int_{S} B_z\Psi \ \textrm{d}^2x.
\end{align}
The most obvious of these is when $S=D_1$,
since $\Psi=0$ everywhere on $\partial D_1$ from the assumption that the side boundary of $V$ is ideal.
Thus, although the work term dominates changes of field line helicity for individual field lines,
its net effect over the entire domain is the same as the integrated effect of 
the typically much smaller voltage drop term.
We conclude that the divergence part of the $\vec{w}\cdot\vec{A}$ term occurs as pairs of opposite polarity,
which redistribute magnetic helicity and are individually typically much stronger than $\Psi$
but which cancel one another in the area integral.  

Area integrals of the voltage drop are also readily interpreted.
Taking the time derivative of Eq.~(\ref{eq:total}) and assuming that $D_1$
and $B_z|_{D_1}$ are fixed in time, the rate of change of total helicity is
\begin{align}
  \frac{\textrm{d}H}{\textrm{d}t} 
  &= \frac{\textrm{d}}{\textrm{d}t}\int_{D_1}B_z\mathcal{A}\ \textrm{d}^2 x
  = \int_{D_1}B_z\frac{\textrm{d}\mathcal{A}}{\textrm{d}t}\ \textrm{d}^2 x.
\end{align}
Then, expanding $\textrm{d}\mathcal{A}/\textrm{d}t$ with Eq.~(\ref{eq:dcurlyA_localized})
and using Eq.~(\ref{eq:area_integrals}) to replace the integral of $B_z\vec{w}\cdot\vec{A}$,
one finds
\begin{align}
  \frac{\textrm{d}H}{\textrm{d}t} 
 &= \int_{D_1}B_z\left(\vec{w}\cdot\vec{A}-\Psi\right)\ \textrm{d}^2 x
 = -2\int_{D_1}B_z\Psi\ \textrm{d}^2 x,
\end{align}
which is equivalent to the volume integral of $-2\vec{E}\cdot\vec{B}$.
In other words, $-2\Psi$ (half from the explicit term in Eq.~(\ref{eq:dcurlyA_localized}) 
and the other half implicit in the work-like term)
represents the net imbalance of helicity sinks and sources along a field line,
which contributes to changing the total helicity 
and is distinct from the redistribution of helicity caused by the divergence part of the $\vec{w}\cdot\vec{A}$ term.
The condition that $\Psi$ is small compared to the work-like term, $\vec{w}\cdot\vec{A}$,
therefore ensures that total helicity is well conserved during the redistribution of helicity by magnetic reconnection.

\begin{figure}
 \includegraphics{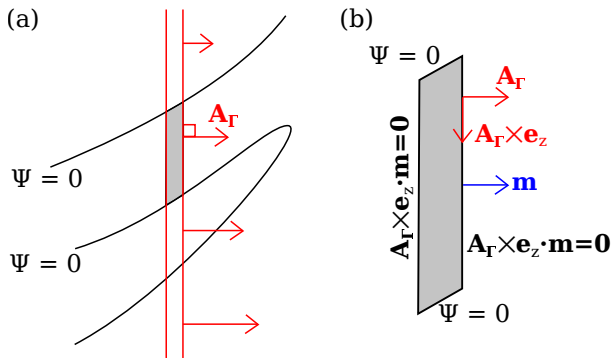}
 \caption{\label{fig:int_areas}Construction of an area inside which 
 positive and negative polarities of $\textrm{d}\mathcal{A}/\textrm{d}t$ cancel.
 (a) shows the curves used to construct the edge of the area: 
 $\Psi=0$ contours (black) and lines perpendicular to the vector potential on the boundary (red).
 (b) shows an enlargement of the shaded area 
 labeled to indicate which quantity is zero on each edge.
 It is readily seen that the line integral of $\Psi \vec{A}\times\vec{e}_z\cdot\vec{m}$ around this shape is zero.}
\end{figure}

Returning to pairing of opposite polarities of $\textrm{d}\mathcal{A}/\textrm{d}t$,
it is also straightforward to construct areas of integration that are smaller than the entirety of $D_1$
and within which the integrated divergence term disappears.
Inspecting Eq.~(\ref{eq:wdotA_int}), the key to this is that contours where $\Psi=0$ 
and curves perpendicular to $\vec{A}_\Gamma$ 
(components of $\vec{A}$ perpendicular to the surface normal of $D_1$) are generally not aligned.
The boundary of a suitable area, $S$, can therefore be built up by 
joining sections of $\Psi=0$ contours (integrand zero since $\Psi=0$)
with sections of curves perpendicular to $\vec{A}_\Gamma$
(integrand zero since $\vec{A}\times\vec{e}_z\cdot\vec{m}=0$).
A sketch is given in Fig.~ \ref{fig:int_areas}.
The spacing between adjacent $\Psi=0$ contours (or points on the same contour) is fixed,
but there is freedom to place the curves perpendicular to $\vec{A}_\Gamma$ arbitrarily close together.
Hence, pairing of increases and decreases of $\mathcal{A}$ 
occurs along every curve perpendicular to $\vec{A}_\Gamma$ 
that connects two $\Psi=0$ contours.
It is therefore seen that redistribution of helicity occurs in a highly organized manner
between specific groups of field lines.

\section{Kinematic Examples}\label{sec:examples}
To verify the properties identified in Section \ref{sec:localized}, 
we now examine a kinematic model of magnetic reconnection 
in a magnetic field with complex connectivity.
The basis of this model is a static magnetic braid
into which a time-dependent ring of magnetic flux is added.

\begin{figure}
 \includegraphics{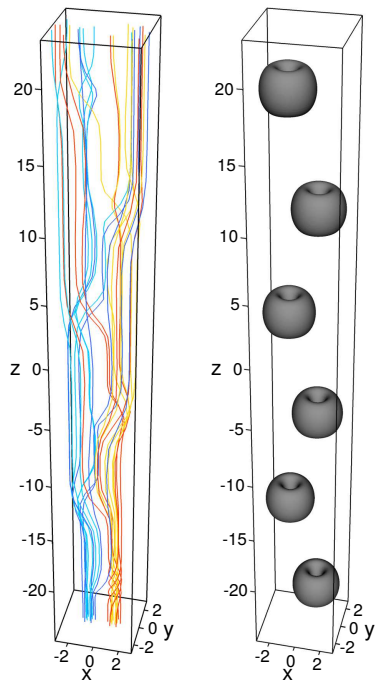}
 \caption{\label{fig:$E^3$}
     Visualization of the $E^3$ magnetic braid, 
     used as the static part of the kinematic model with magnetic complexity.
     (left) a selection of magnetic field lines which show the braided nature of the field.
     (right) isosurfaces of horizontal magnetic field strength, 
     revealing the magnetic flux rings that are superimposed on a uniform vertical magnetic field.  
     Flux rings centered on $x=1$ are right-handed and flux rings centered on $x=-1$ are left-handed.
 }
\end{figure}

For the static component of the magnetic field, 
we use the $E^3$ magnetic braid detailed by \citet{2009WilmotSmith},
which consists of a superposition of three left-handed and three right-handed rings 
of horizontal magnetic flux with a vertical uniform magnetic field.
This braid is visualized in Fig.~ \ref{fig:$E^3$}
and attention is drawn to the complex field line mapping.
It was chosen only for convenience and any other sufficiently 
complex braid would serve just as well.  

Field line helicity is calculated subject to the gauge constraint
${\vec{A}\times\vec{n}|_{\partial V}=\vec{A}^\textrm{ref}\times\vec{n}|_{\partial V}}$, 
where $\vec{n}$ is the outward normal on $\partial V$, and 
\begin{align}
  \vec{A}^\textrm{ref} &= \frac{B_0}{2}\left(-y\vec{e}_x+x\vec{e}_y\right),\label{eq:work_simplified_by_gauge}
\end{align}
which corresponds to a uniform vertical reference magnetic field, $\vec{B}^\textrm{ref}=B_0\vec{e}_z$.
This particular constraint ensures that helicities are the same as for 
the winding gauge of \citet{2014PriorYeates}, 
which measures winding in an untwisted frame. 
It also gives ${H=H_R}$, which is zero for the $E^3$ braid.

\begin{figure}
 \includegraphics{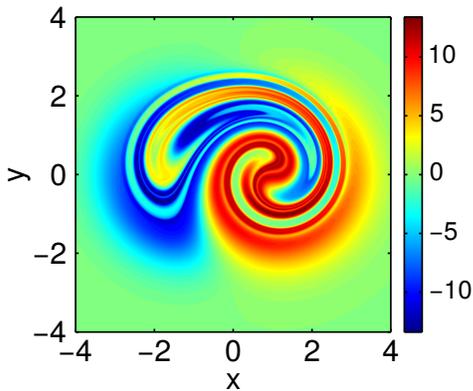}
 \caption{\label{fig:$E^3$_curlyA}
     Field line helicity, $\mathcal{A}$, for the $E^3$ magnetic braid, mapped on the $D_0$ surface, $z=-24$.
     Small scales are evident and come from the complexity of the field line mapping.
 }
\end{figure}

The map of the initial field line helicity on $D_0$ is shown in {Fig.~ \ref{fig:$E^3$_curlyA}}.
It exhibits considerable complexity, 
with scales much shorter than those in the magnetic field ({Fig.~ \ref{fig:$E^3$}}) 
coming from the complexity of the field line mapping.

Since our gauge restriction fixes the components of $\vec{A}$ tangent 
to the domain boundary and since $\vec{E}=0$ on $\partial V$ 
when reconnection is localized within the domain, Eq.~(\ref{eq:faraday_A}) implies that 
$\vec{n}\times\nabla\Phi|_{\partial V}=0$, i.e. ${\Phi=\textrm{const}}$ on $\partial V$.
Thus, $\Phi(\vec{x}_0)=\Phi(\vec{x}_1)$ for any field line,
which makes the gauge terms cancel in Eq.~(\ref{eq:dcurlyA}).
Calculation of $\vec{w}\cdot\vec{A}$ terms is simplified too
because defining $\vec{w}\cdot\vec{n}=0$ on $D_0$ and $D_1$
gives $\vec{w}\cdot\vec{A}=\vec{w}\cdot\vec{A}^\textrm{ref}$ at field line endpoints.
When reconnection is localized within the domain, this simplifies further to give
\begin{align}\label{eq:wdotA_example}
 \vec{w}\cdot\vec{A} &= \frac{r}{2}\frac{\partial \Psi}{\partial r}
\end{align}
where $r$ is the radial distance from $(x,y)=(0,0)$.

Attention is drawn to the fact that provided the boundary condition on $\vec{A}$
is satisfied, $\mathcal{A}$ and $\textrm{d}\mathcal{A}/\textrm{d}t$ do not change under
the allowable gauge transformations.
This can be attributed to $\mathcal{A}$ obtaining a physical meaning
as a measure of winding with reference to a fixed frame
or as a density of relative helicity,
due to the boundary condition on $\vec{A}$.

Time-dependence is imposed via a prescribed electric field,
\begin{align}\label{eq:E}
  \vec{E} &= -B_0ak\exp\left(-\frac{x^2}{a^2}-\frac{(y-1)^2}{a^2}-\frac{z^2}{L^2}\right)\vec{e}_z,
\end{align}
which corresponds to growth of a time-dependent flux ring
at the midplane with its center offset slightly from the central axis
(readily confirmed by taking the curl of Eq.~(\ref{eq:E})).
The horizontal flux added over time
changes magnetic connectivities between $D_0$ and $D_1$,
and the kinematic model acts as a proxy for magnetic reconnection in this regard.
Parameters were set as $B_0=1$, $k=0.5$, $L=2$ and $a=\sqrt{2}/4$,
giving a time-dependent flux ring spatially smaller than the static ones,
thus the model describes reconnection that is localized 
to a small region within the larger field.

Compared to other models of reconnection,
kinematic models have the limitation that the electric field is prescribed
rather than arising self-consistently from an algebraic Ohm's law.
The model described is nonetheless sufficient to verify the properties
identified in Sec.~\ref{sec:localized}, 
which do not rely on any specific form for Ohm's law.
We have also obtained similar results using resistive MHD simulations,
although the kinematic model gives a clearer illustration
because it limits reconnection in the complex magnetic field to a single location.

The rate of change of field line helicity for the 
kinematic model was computed as follows.
First, we set $\Psi=0$ on $D_0$,
which also gives $\vec{w}=0$ there by Eq.~(\ref{eq:w_boundaries}).
Thus, the full evolution equation (\ref{eq:dcurlyA})
reduces to the simplified form given by Eq.~(\ref{eq:dcurlyA_localized}).
The values of $\Psi$ on $D_1$ were obtained by integrating $E_{||}$
along the magnetic field, in keeping with its definition in Eq.~(\ref{eq:psi})
and using the boundary condition $\Psi|_{D_0}=0$ to fix the constant of integration.
Finally, the $\vec{w}\cdot\vec{A}$ term on $D_1$ was evaluated  using
Eq.~(\ref{eq:wdotA_example}).
Note that the approach of evaluating $\Psi$ by integration of
$E_{||}$ along field lines and $\vec{w}$ on the boundary from $\nabla \Psi$
works not only when Ohm's law is specified explicitly (e.g. Eq.~(\ref{eq:ohms}))
but also when it arises implicitly from a prescribed electric field as in the kinematic model.

\begin{figure}
 \includegraphics{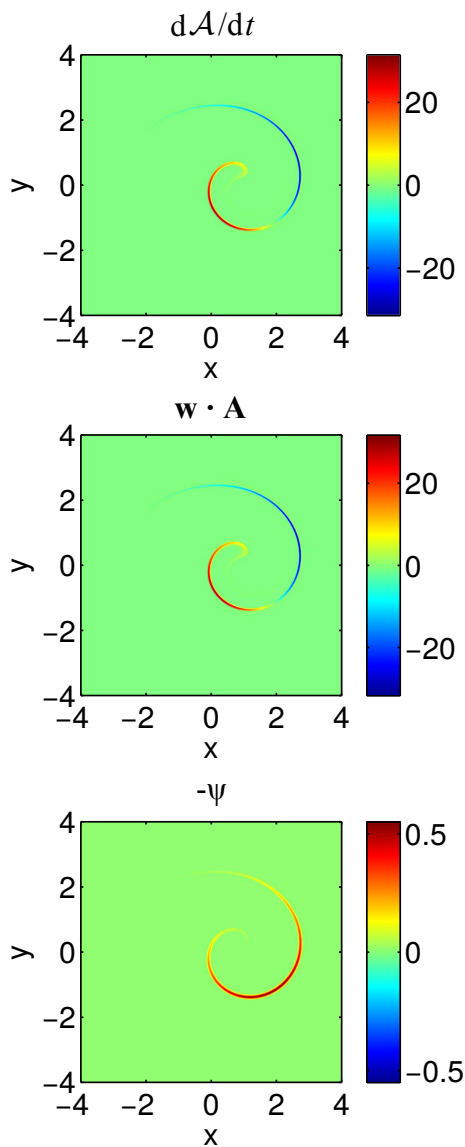}
 \caption{\label{fig:kin_$E^3$}
     Evolution of field line helicity on $D_0$ for the kinematic model 
     with magnetic complexity.
     (top) Map of $\textrm{d}\mathcal{A}/\textrm{d}t$ (extreme value of 31.4).
     (middle) Map of the $\vec{w}(\vec{x}_1)\cdot\vec{A}(\vec{x}_1)$ term (extreme value of 31.7).
     (bottom) Map of the $-\Psi(\vec{x}_1)$ term (extreme value of 0.55).
     Quantities are plotted on $D_0$ where field lines are fixed to stationary plasma;
     the evolution equation terms were evaluated at the field line's conjugate point on $D_1$, $\vec{x}_1 = F(x,y)$, and mapped back to $D_0$.
 }
\end{figure}

Figure \ref{fig:kin_$E^3$} shows $\textrm{d}\mathcal{A}/\textrm{d}t$, 
and the contributions from the $-\Psi(\vec{x}_1)$ and 
$\vec{w}(\vec{x}_1)\cdot\vec{A}(\vec{x}_1)$ terms.
All quantities are plotted mapped to $D_0$ where field lines are fixed 
to the stationary plasma by the choice $\Psi=0$ on $D_0$,
i.e. $\textrm{d}\mathcal{A}/\textrm{d}t\equiv \partial \mathcal{A}/\partial t$ on $D_0$.
It is immediately apparent that $\textrm{d}\mathcal{A}/\textrm{d}t$ 
is dominated by the $\vec{w}\cdot\vec{A}$ term,
which has an extreme value (31.7) that is 57.6 times greater 
than the extreme value of the $-\Psi$ term (0.55).
This finding is in good agreement with the analytic results of Sec.~\ref{sec:dominant}, 
which argued that the work-like term should be dominant given a 
complex magnetic field.

\begin{figure}
 \includegraphics{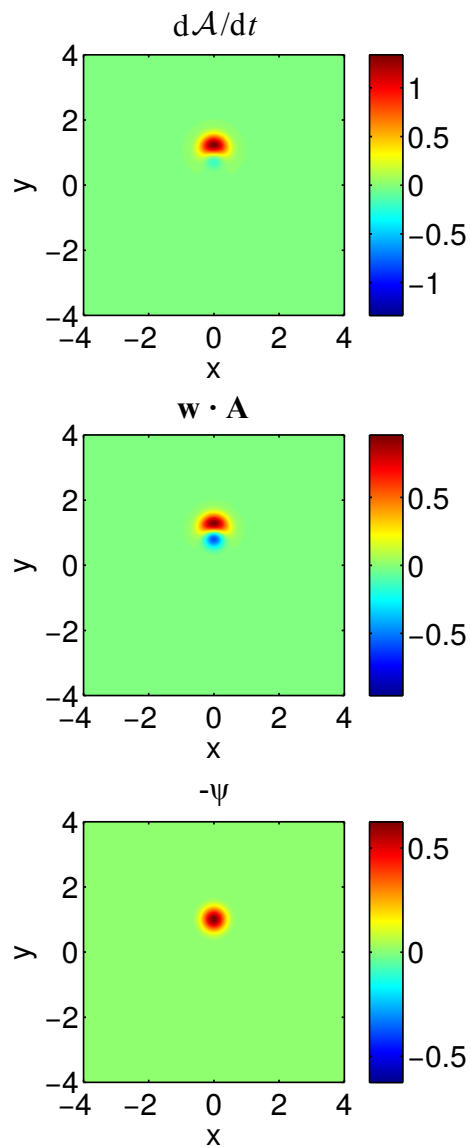}
 \caption{\label{fig:kin_nobraid}
     Evolution of field line helicity on $D_0$ for the kinematic model 
     without magnetic complexity,
     in the same format as Fig.~~\ref{fig:kin_$E^3$}.
     The extreme values are: 1.34 (${\textrm{d}\mathcal{A}/\textrm{d}t}$, top panel), 
     0.96 (${\vec{w}(\vec{x}_1)\cdot\vec{A}(\vec{x}_1)}$, middle panel) and 
     0.63 (${-\Psi(\vec{x}_1)}$, bottom panel).
     }
\end{figure}

The importance of complexity in the field line mapping
is demonstrated if the static braid 
is replaced with a straight uniform magnetic field
(for which $\mathcal{A}=0$ everywhere), 
in which case the results shown in Fig.~ \ref{fig:kin_nobraid} are obtained.
With the removal of braiding, the extreme values of the $\vec{w}\cdot\vec{A}$ 
and $-\Psi$ terms become 0.96 and 0.63 respectively,
giving a ratio of 1.5, i.e. neither term dominates strongly when the field lacks complexity.

\begin{figure}
 \includegraphics{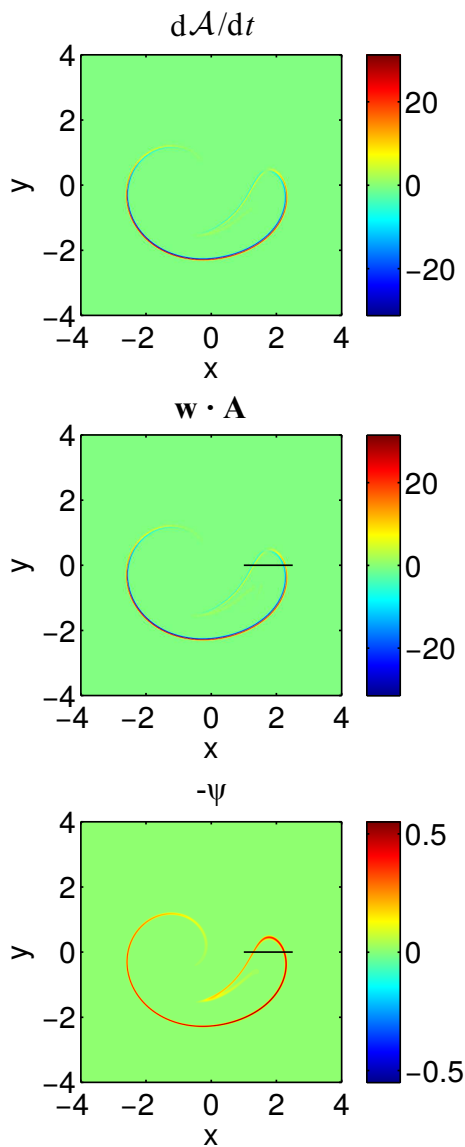}
 \caption{\label{fig:pairing}
     Pairing of $\vec{w}\cdot\vec{A}$ polarities (hence polarities of $\textrm{d}\mathcal{A}/\textrm{d}t$) 
     on $D_1$ for the kinematic model with magnetic complexity.
     The horizontal black line superimposed on the middle and bottom panels shows the line 
     on which quantities are inspected in Fig.~ \ref{fig:lineout}.
     Analytic theory predicts pairs of positive and negative polarities where 
     radial lines on $D_1$ (which are perpendicular to the vector potential)
     cross the threadlike region where $\Psi\neq0$. 
    The prediction holds very well.
     }
\end{figure}

\begin{figure}
 \includegraphics{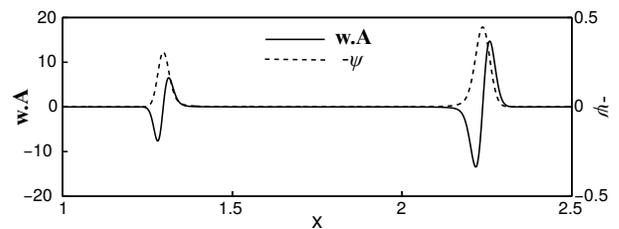}
 \caption{\label{fig:lineout}
     Pairing of $\vec{w}\cdot\vec{A}$ on $D_1$, along a line perpendicular to $\vec{A}^\textrm{ref}$
     (the section of the line $y=0$ shown in Fig.~ \ref{fig:pairing}).
     The solid curve plots $\vec{w}\cdot\vec{A}$ (left axis) which has a bipolar character,
     and the dashed curve shows $-\Psi$ (right axis) which is monopolar.
     The sign of $\vec{w}\cdot\vec{A}$ changes when the direction of field line motion reverses.
     }
\end{figure}

The kinematic example with magnetic complexity (Fig.~ \ref{fig:kin_$E^3$}) 
also shows pairing of opposite polarities of $\textrm{d}\mathcal{A}/\textrm{d}t$,
such that field line helicity is redistributed by the work-like term,
while the total helicity changes much more slowly.
This too confirms our expectation from analytic results (Sec.~\ref{sec:pairing}).
The analysis in Sec.~\ref{sec:pairing} predicts that pairing of opposite polarities is well organized.
In particular, when terms are examined on $D_1$,
we expect to see pairing along lines perpendicular to the vector potential that connect two $\Psi=0$ contours.
This property is investigated in Figs. \ref{fig:pairing} and \ref{fig:lineout}.
For our choice of gauge, curves on $D_1$ perpendicular to $\vec{A}$ are simply radial lines.
Moreover, the kinematic example gives a profile of $\Psi$ which is very localized on $D_1$, 
$\Psi\neq0$ being restricted to a threadlike structure.
The theory therefore predicts that $\vec{w}\cdot\vec{A}$ (and hence $\textrm{d}\mathcal{A}/\textrm{d}t$) 
is organized in pairs of opposite polarity where radial lines cross the threadlike region where $\Psi\neq0$.
Consulting Figs. \ref{fig:pairing} and \ref{fig:lineout}, this expectation is confirmed.

Note that pairing of polarities on $D_1$ implies pairing on $D_0$ via the field line mapping.
This is apparent when comparing Figs. \ref{fig:kin_$E^3$} and \ref{fig:pairing}.
$B_z=1$ on each boundary, and all quantities shown are defined for each field line, hence the figures are area-preserving rearrangements of one-another.
However, despite the simple structure of pairing on $D_1$ along radial lines,
the complexity of the field line mapping means that 
the corresponding structure of pairing on $D_0$ 
generally does not have an obvious analogous form.

Returning to the topic of pairing of positive and negative polarities on the top boundary, 
an intuitive feel for the exact analytic result may be gained as follows.
Field lines are fixed to the fluid on $D_0$ by our choice that $\Psi=0$ there, 
but non-ideal effects in the domain allow them to slip on $D_1$.
If dynamics are localized inside the domain, then it follows from 
Eq.~(\ref{eq:w_boundaries})
that motion of field line endpoints on $D_1$ at any moment is along contours of $\Psi$
(clockwise around maxima of $\Psi$ and counterclockwise around minima of $\Psi$).
Now consider a curve on $D_1$ perpendicular to the vector potential, 
parameterized by distance $l$ along it, which connects two points where $\Psi=0$.
In our example, the radial line considered in Fig \ref{fig:lineout} is one such line.
Where $\textrm{d}\Psi/\textrm{d}l>0$, $\vec{w}$ will have one sign 
and where $\textrm{d}\Psi/\textrm{d}l<0$ it will have the other sign.
Moreover, the flows in the opposite directions balance 
because $\Psi$ is continuous and the magnitude of $\vec{w}$ is proportional to the gradient.
The final assumption to transfer this balance of field line motions to the helicity equation 
is to assume that the width of the $\Psi\neq0$ region
is much smaller than the length scale over which $\vec{A}^\textrm{ref}$ changes appreciably.
Then, $\vec{A}^\textrm{ref}$ may be treated as constant to leading order along the curve of interest, 
and it is seen that the oppositely directed field line motions 
correspond to opposite polarities of $\vec{w}\cdot\vec{A}$
(for our example, also refer to Eq.~(\ref{eq:work_simplified_by_gauge})).
This discussion is less rigorous than the analytic presentation of pairing in Sec.~\ref{sec:pairing},
but it may help to develop a feeling for the underlying physics.

Finally, it is noted that the introduction of an additional magnetic flux ring 
means that the total helicity of the field is changing.
This happens self-consistently through the helicity source $-2\vec{E}\cdot\vec{B}$.
We have already suggested, however, that if the preexisting magnetic field 
has a complex field line mapping, 
then the field line helicity is rearranged on a shorter timescale than that 
associated with changes to the total helicity.
This is confirmed by computing the flux integral of $\textrm{d}\mathcal{A}/\textrm{d}t$ 
and the flux integral of its unsigned value.
The results are that $\textrm{d}H/\textrm{d}t=0.5$,
while the integral of $B_z|\textrm{d}\mathcal{A}/\textrm{d}t|$ over the domain is 11.4.
The latter integral is 22.8 times greater than the former and 
is very strongly dominated by the $\vec{w}\cdot\vec{A}$ term.
Thus, it is clear that reconnection in a complex magnetic field acts primarily
to redistribute the field line helicity, 
and that this occurs much more rapidly than the total helicity changes.
(When there is no preexisting pattern of $\mathcal{A}$ to be redistributed, 
which is the case for the example shown in Fig.~ \ref{fig:kin_nobraid}, 
then the flux integrals are more similar.)

\section{Summary \& Conclusions}\label{sec:conc}

This paper has shown how an evolution equation can be derived for field line helicity,
$\mathcal{A}$, and examined its properties.
For a suitable restriction of the gauge, 
$\mathcal{A}$ changes only if the connectivity of the field changes,
either via plasma motion on the domain boundary 
or magnetic reconnection within the domain.
Then, the evolution equation becomes the sum of two terms:  a work-like term, 
namely the scalar product of the generalized field line velocity with the vector potential 
at field line end points, and the voltage drop along the field line.

Localized magnetic reconnection within a complex magnetic field
was given particular consideration.
This is relevant, for example, for magnetic relaxation of complex magnetic fields.
Using the evolution equation for $\mathcal{A}$, simple scaling arguments show that 
evolution of magnetic helicity is strongly dominated by the work-like term under these conditions.
Moreover, the work-like term has a property that means changes to field line helicity occur in 
paired regions of opposite polarity, which approximately cancel one another overall.
It follows that magnetic reconnection in a complex magnetic field
serves primarily to rearrange helicity, and has a relatively small impact on the total helicity.

The idea that total helicity is approximately conserved on the timescale 
during which magnetic reconnection rearranges an initially complex magnetic field is longstanding.
Notably, it is the conjecture underlying Taylor relaxation \citep{1974Taylor,1986Taylor}
and has previously been justified on the basis that an Cauchy-Schwarz inequality 
implies that helicity decay occurs on a diffusive time scale that is longer than the time over which 
energies change, especially when reconnection occurs in a small proportion of the total volume \citep{1984Berger}. 
For turbulent situations, the inverse cascade of magnetic helicity to large scales\citep{1975Frisch},
where dissipation is inefficient, contrasts with the direct cascade of energy to small scales,
and this has also been used to argue for approximate helicity conservation.
This paper provides an alternative and complementary justification for 
helicity conservation during localized magnetic reconnection,
with the advantages that it explicitly shows the rapidity of helicity reorganization
and provides new detail about its underlying mechanics.
Our results also emphasize the importance of existing magnetic complexity, 
without which the timescales of reorganization and net helicity change are separated at most weakly.

Now that the evolution of field line helicity is better understood,
it should be possible to refine our understanding of relaxation via magnetic reconnection.
For example, it is not fully accurate to say that approximate
conservation of total helicity is the only
helicity constraint (the origin of Taylor's relaxation hypothesis).
Rather, field line helicity is reorganized in a manner 
prescribed by the dominant terms in the evolution equation.
The evolution equation may therefore reveal new 
constraints on magnetic relaxation via reconnection.
That possibility will be the subject of future work.

\begin{acknowledgments}
  This work was supported by the Science and Technology Facilities Council (UK) 
  through consortium grants ST/K000993/1 and ST/K001043 
  to the University of Dundee and Durham University.
  We thank an anonymous referee for helpful comments that improved the manuscript.
\end{acknowledgments}

\bibliography{curlyAevolution}

\end{document}